\documentclass[preprint,prc,aps,eqsecnum,epsf,amssymb]{revtex4}
\usepackage{graphicx}% Include figure files
\newcommand{\qq}{{|\mbox{\boldmath $q$}|}}

\begin{document}
\preprint {WIS-04/26 Dec-DPP}
%\draft
\date{\today}
\title{On Distribution Functions for Partons in Nuclei}
\author{A.S. Rinat and M.F Taragin}
\address{Weizmann Institute of Science, Department of Particle Physics,
Rehovot 76100, Israel}
%\maketitle
\begin{abstract}

We suggest that a previously conjectured relation between Structure Functions 
(SF) for nuclei and nucleons also links distribution functions (df) 
for partons in  a nucleus and in nucleons. The above suggestion ensures in 
principle identical results for SF $F_2^A$, whether computed with hadronic 
or partonic degrees of freedom. In practice there are differences, 
due to different $F_2^n$ input. We show that the thus defined nuclear 
parton distribution functions (pdf) respect standard sumrules. In addition 
we numerically compare some moments of nuclear SF, and find 
agreement between results, using hadronic and partonic descriptions. We 
present computations of EMC ratios for both and compare 
those with hadronic predictions and data. In spite of substantial 
differences in the participating SF, the two representations produce 
approximately the same EMC ratios. The apparent correlation between the
above deviations is ascribed to a sumrule for $F_2^A$. We conclude with a 
discussion of alternative approaches to nuclear pdf.

\end{abstract}

We suggest that a previously conjectured relation between Structure
Functions
(SF) for nuclei and nucleons also links distribution functions (df)
for partons in  a nucleus and in nucleons. The above suggestion ensures in
principle identical results for SF $F_2^A$, whether computed with hadronic
or partonic degrees of freedom. In practice there are differences,
due to different $F_2^n$ input. We show that the thus defined nuclear
parton distribution functions (pdf) respect standard sumrules. In addition
we numerically compare some moments of nuclear SF, and find
agreement between results, using hadronic and partonic descriptions. We
present computations of EMC ratios for both and compare
those with hadronic predictions and data. In spite of substantial
differences in the participating SF, the two representations produce
approximately the same EMC ratios. The apparent correlation between the
above deviations is ascribed to a sumrule for $F_2^A$. We conclude with a
discussion of alternative approaches to nuclear pdf.

\maketitle

\section{Introduction.}

A large body of data is presently available on inclusive lepton scattering 
from nuclei and on subsequently extracted nuclear Structure Functions (SF).
Standard treatments of those data use hadron degrees of freedom, whereas in
an alternative approach, one employs df of partons in nuclei. In one of the 
first papers on EMC ratios, Akulinichev $et\,al$ related df for quarks in a 
nucleon, nucleons in nuclei and quarks in nuclei by a generalized convolution, 
schematically written as \cite{akul} (see also Refs. \cite{bc,bbm}) 
\begin{eqnarray}
f_{q/A}=f_{N/A}*f_{q/N},
\label{a1}
\end{eqnarray}
Variations of the above and various approximations have since been the 
preferred tool for the analysis of EMC ratios, whether in terms of the 
hadron or a parton representation. 

There exist alternative treatments of nuclear pdf. For instance Frankfurt 
and Strikman discuss nuclear pdf, but do not relate those to the pdf of a 
nucleon \cite{fs}. Other approaches parametrize information on SF at some
scale $Q_0^2$, and then evolve to a desired $Q^2$. Data then determine the 
introduced 
parameters \cite{esk,kum}. Next we mention quark models for nuclei, which 
have been used in direct calculations of df. Their potential is presumably 
limited to the lightest nuclei \cite{benesh}. Finally there are 
approaches, where the effect of a nuclear medium on a nucleon or a quark 
is replaced by mean fields \cite{mineo,stef,ss}. 

In the present note we make 
a simple, nearly natural choice for nuclear pdf, which is free of adjustable 
parameters. Those satisfy basic sum rules and produce the same $F_2^A$ as 
computed from a hadronic base. In the conclusion we compare some of the 
above-mentioned alternative proposals with our choice. 

\section {A few essentails.}

We start with the cross section per nucleon for the scattering 
of unpolarized electrons with energy $E$ over an angle $\theta$
\begin{eqnarray}
\frac{d^2\sigma^A(E;\theta,\nu)}{d\Omega\,d\nu}
=\sigma_M(E;\theta,\nu)\bigg\lbrack\frac {2xM}{Q^2}
  F_2^A(x,Q^2)+ \frac{2}{M}F_1^A(x,Q^2){\rm tan}^2(\theta/2) \bigg\rbrack
\label{a2}
\end{eqnarray}
$\sigma_M$ is the Mott cross section and $F_{1,2}^A(x,Q^2)$ are the 
standard nuclear SFs per nucleon. Those depend on the squared 4-momentum 
transfer $q^2=-Q^2=-(\qq^2-\nu^2)$ and on the Bjorken variable 
$0 \le x=Q^2/2M\nu \le A$ in terms of the nucleon mass $M$. 

Next we make explicit the specific relation (\ref{a1}) between nuclear and 
nucleonic SF \cite{gr}
\begin{mathletters}  
\label{a3}
\begin{eqnarray}               
  F^A_k(x,Q^2)&=&\int_x^A\frac {dz}{z^{2-k}} f^{PN,A}(z,Q^2) 
     F_k^{\langle N \rangle}\bigg (\frac {x}{z},Q^2\bigg )
\label{a3a}\\
  F_k^{\langle N \rangle}=\bigg [ZF_k^p +NF_k^n \bigg ]/2A
&=& \bigg [1-\frac {\delta N}{A}\bigg]F_k^p+
\bigg [1+\frac {\delta N}{A}\bigg ]F_k^n
\label{a3b}
\end{eqnarray}
\end{mathletters}
$F_k^{p,n}$ are SFs of $p,n$, whereas $F_k^{\langle N \rangle}$ 
defines a nucleon SF, obtained by weighting $F_k^{p,n}$ with $Z,N$. 
$\delta N/A$ denotes the relative neutron excess. 

Our approach draws on the Gersch-Rodriguez-Smith (GRS) theory for inclusive 
scattering of non-relativistic projectiles \cite{grs}, in which case the 
linking $f^{PN,A}$  is the SF of a fictitious nucleus, composed of point-like 
nucleons. Eq. (\ref{a3a}) results from a covariant generalization of the 
GRS theory \cite{gr1}. The theory is non-perturbative with on-mass shell
nucleon SFs. 

Eq. (\ref{a1}) has originally been formulated in the Bjorken limit
\cite{akul}. We have postulated (\ref{a3a}) to hold for finite $Q^2$ 
\cite{gr} and an extensive body of data in the ranges $x\gtrsim 0.2\,
;Q^2\gtrsim 2.5\,$GeV$^2$ \cite{rtval,rtv} appears to be accounted for by 
the relation (\ref{a3a}) \cite{rt2,rt1,vkr}.

Eq. (\ref{a3a}) does not contain a term, which account for quarks in 
virtual mesons \cite{lls}, and it also lacks  (anti-)screening effects 
\cite{wise}. Those are negligible for $x\gtrsim 0.2$ and we limit ourselves 
to that range \footnote{ We also disregard mixing of nucleonic SF, the 
relative importance of which diminishes with increasing $Q^2$ \cite{atw}.}. 

In the following it appears useful to separate the nucleon SF 
$F_k^N=F_k^{N,NE}+F_k^{N,NI}$ into nucleon elastic and inelastic components, 
which correspond to absorption processes of a virtual photon on a $N$, 
$\gamma^*+N\to N$, (NE), or $\gamma^*+ N\to$ (hadrons,partons) (NI). Elastic 
components for a $N$ are proportional to the standard combinations of squared 
electro-magnetic form factors $G_{E,M}^N(Q^2)$ and vanish, unless $x=1$. With 
$[{\tilde G}^{\langle N \rangle}]^2=[Z(G^p)^2+N(G^n)^2]/A$, one has
\begin{mathletters}
\label{a4}
\begin{eqnarray}
F_1^{N,NE}(x,Q^2)&=&\frac{1}{2}\delta(1-x)
[{\tilde G}_M^{\langle N \rangle}(Q^2)]^2
\label{a4a}\\
F_2^{N,NE}(x,Q^2)&=&\delta(1-x)
\frac {[{\tilde G}_E^{\langle N \rangle}(Q^2)]^2
+\eta  [{\tilde G}_M^{\langle N \rangle}(Q^2)]^2}{1+\eta}
\label{a4b}
\end{eqnarray}
\end{mathletters}
The corresponding nuclear NE (QE) components are from Eq. (\ref{a3a}) seen 
to be
\begin{mathletters}
\label{a5}
\begin{eqnarray}
 F_1^{A,NE}(x,Q^2)&=&\frac {f^{PN,A}(x,Q^2)}{2}
[{\tilde G}_M^{\langle N \rangle}(Q^2)]^2] \,
  \label{a5a}\\
\noalign{\medskip}
  F_2^{A,NE}(x,Q^2)&=&xf^{PN,A}(x,Q^2) 
\frac{[{\tilde G}_E^{\langle N \rangle}(Q^2)]^2+ 
\eta  [{\tilde G}_M^{\langle N \rangle}(Q^2)]^2}{1+\eta}
  \label{a5b}
\end{eqnarray}
\end{mathletters}
In particular for the lightest nuclei the normalized $f^{PN,A}$ peaks around 
$x\approx 1$, and the same holds for the above QE components  
$F_k^{A,NE}(x,Q^2)$. The above summarizes elements of a hadronic description 
of nuclear SF: we now turn to a partonic representation.

\section{A simple choice for nuclear parton distribution functions.}

We start with the leading order twist contributions to the dominant NI 
components of nucleon SF for finite $Q^2$. With no danger of confusion we 
shall omit one or both arguments $x,Q^2$. Neglecting heavy quark 
contributions and decomposing quark df $q=q_v+{\bar q}$ into valence
and sea quarks parts, one has 
\begin{eqnarray}
F_2^p&=&\frac {x}{9}\bigg (4u_v +d_v+8{\bar u}+ 2{\bar d}+2s\bigg )
\nonumber\\
F_2^n&=&\frac {x}{9}\bigg (u_v +4d_v+2{\bar u}+ 8{\bar d}+2s\bigg )
\label {a6}
\end{eqnarray}
Similarly for a $'$nucleon$'$, defined as the $Z,N$ weighted $p,n$, one has
\begin{eqnarray}
F_2^{\langle N \rangle}\equiv \sum_ia_ixq_i=\frac {5x}{18}\bigg [u_v+d_v+ 
2{\bar u}+2{\bar d}+\frac {4}{5}s-\frac{3\delta N}{5A}
(u_v-d_v+2{\bar u}-2{\bar d})\bigg ],
\label{a7}                                         
\end{eqnarray}
Nuclear pdf ought to reproduce nuclear SF $F_k^A$, just as proton pdf do for 
$F_k^p$, Eq. ({\ref{a6}), but that necessary requirement is insufficient
for their determination. In a construction we proceed in two steps. 
First we choose $F_2^A(q^A)$ to be the same combination of df of partons 
in a nucleus, as the above $F_2^{\langle N \rangle}$, Eq. (\ref{a7}), 
is for the averaged nucleon 
\begin{eqnarray}                                      
F_2^A ={\sum_i}a_i xq_i^A                           
&=&\frac{5x}{18}\bigg [u_v^A+d_v^A+                   
2{\bar u}^A+2{\bar d}^A+\frac{4}{5}s^A-\frac{3\delta N}{5A}
(u_v^A-d_v^A+2{\bar u}^A-2{\bar d}^A)\bigg ]
\label{a8}
\end{eqnarray}
Upon substitution into Eq. (\ref{a3a}), and using  Eq. (\ref{a7}), one finds 
\begin{eqnarray}
F_2^A={\sum_i}a_i \bigg [f^{PN,A}x*q_i^A\bigg ]
\label{a9}
\end{eqnarray}
Comparison with Eq. (\ref{a8}) still does not result in a unique expressions 
for each individual pdf $q_i^A$. Then guided by Eq. (\ref{a3a}), we make the 
following second choice, with no flavor mixing of valence and sea quarks
\begin{eqnarray}
xq_i^A(x,Q^2) &\equiv & \int_x^A dz f^{PN,A}(z,Q^2)\bigg (\frac{x}{z}\bigg )
u_v\bigg (\frac {x}{z},Q^2 \bigg )
\nonumber\\
x{\bar q_i}^A(x,Q^2) &\equiv & \int_x^A dz f^{PN,A}(z,Q^2)\bigg (\frac {x}{z}
\bigg )d_v\bigg (\frac {x}{z},Q^2 \bigg )
\nonumber\\
xg^A(x,Q^2) &\equiv & \int_x^A dz f^{PN,A}(z,Q^2)\bigg (\frac{x}{z}\bigg )
g\bigg (\frac {x}{z},Q^2 \bigg )
\label{a10}
\end{eqnarray}
Eqs. (\ref{a10}) with one $f$ for all partons (in case for the PWIA) had 
already be suggested by Berger and Coester \cite{bc}. 

In view of the meager 
experimental information on non-valence parton distributions in nuclei, we 
shall also investigate changes when non-valence df are not affected 
by the nuclear medium (cf. \cite{esk,kum}), thus
\begin{eqnarray}
{\bar q}^A\equiv{\bar q};\, s^A={\bar s}^A=s;\, g^A=g^N, 
\label{a11}
\end{eqnarray}
Eqs. (\ref{a8}), (\ref{a10}) (or alternatively Eq. (\ref{a11})) manifestly 
produce the same $F_2^A$ in the parton and the hadronic representation, 
provided one uses exactly the same input $f^{PN,A}$ and 
$F_2^{\langle N \rangle}$ in both.
 
In practice this is not the case, in particular not 
for $F_2^n$ in $F_2^{\langle N \rangle}$. In the absence of direct 
information, the CteQ parametrizations exploit data on $F_2^D$. Using the 
$'$primitive$'$ approximation $F_2^n(x)=2F_2^D(x)-F_2^p$, $\,\,\,F_2^n(x)$ 
replaces $F_2^D$ as input. However, the above approximation deteriorates 
with $x$, increasing from $x\gtrsim (0.25-0.30)$. As we shall shortly 
demonstrate, its use for larger $x$ leads to misfits with data.

As to the options in the version CteQ6 \cite{cteq} we selected the one, 
with $F_2^p$, closest to the Arneodo parametrization of 
resonance-averaged data \cite{arneo}. This appears possible for 
$x\lesssim 0.6-0.7$. Using $SU_3$ symmetry one obtains in the usual way 
$F_2^n$ in terms of $q_i, {\bar q_i}$.

In contrast, in the hadronic approach one stays as close as possible to 
data. For not too high $Q^2$, one can exploit parametrizations of actual 
$F_2^p$ data \cite{christy}. However, the range $Q^2\gtrsim 3.5\,$GeV$^2$ of 
our interest borders the limits of validity of the data parametrizations, 
we chose to use the above-mentioned Arneodo representation of $F_2^p$ 
\cite{arneo}. As to $F_2^n$, it has been obtained for $all\,x$ by an indirect 
extraction method (see for instance Ref. \cite{rtn}). Fig. 1 for 
$C=F_2^n/F_2^p$ in the two 
representations, $Q^2=3.5, 5.0\,$GeV$^2$, makes manifest the above-mentioned 
differences, which increase with $x$ \footnote{ Lacking experimental 
information on input elements (e.g. $G_E^n$) for higher $Q^2$, we occasionally 
use extrapolations from lower $Q^2$ \cite{bba}.}. 

A second subtle difference between the representations is the validity of 
Eq. (\ref{a3a}) and (\ref{a10}). The former one is explicitly limited to the 
nucleonic part of $F_2^A$, i.e. roughly for $x\gtrsim 0.2$, while the latter 
is conceivably valid out to lower $x$ (cf. Ref. \onlinecite{armesto}). 

The above described freedom in the choice of df for partons in nuclei is not
unlimited, since df are constrained by sum rules. We first check those, which 
are directly related to the normalization of $f^{PN,A}$:

a) For any linear combination $C$ of df for valence quarks 
\begin{mathletters}
\label{a12}
\begin{eqnarray}
\int_0^A dxC^A&=&\int_0^1 dxC^N
\label{a12a}\\
\int_0^A dx (u_v^A+d_v^A)&=& \int_0^1 dx (u_v+d_v)=3
\label{a12b}\\
\int_0^A dx \frac {2u_v^A+d_v^A}{3}&=& \int_0^1 dx \frac{2u_v+d_v}{3}=1
\label{a12c}
\end{eqnarray}
\end{mathletters}
Eqs. (\ref{a12a}), (\ref{a12b}) are for $C^A=u_v^A+d_v^A=u^A-{\bar u}^A+d^A-
{\bar d}^A$, respectively $C^A=(2u_v^A-d_v^A)/3$. Those express the 
conservation (per nucleon) of the number of valence quarks in nuclei (baryon 
number) and of charge. 

b) For any linear combination $C=\alpha_u u_v+\alpha_d d_v$ of nuclear pdf  
\begin{eqnarray}
C^A(0,Q^2)&=&C^N(0,Q^2)
\nonumber\\
C^A(x_0,Q^2)&\approx &C^N(x_0,Q^2)  
\label{a13}
\end{eqnarray}
By construction, Eq. (\ref{a3a}) holds  only for the contribution of partons 
from nucleons, and somehow the same is the case for (\ref{a10}). For those 
parts one proves the above equality for $x=0$, whereas for $0.18\gtrsim x_0
\gtrsim 0$, Eq. (\ref{a13}) is an approximation.

c) For any combination $xC(x)=x\sum_i \alpha_i q_i(x,)$ or $x\sum_i \alpha_i
{\bar q_i}(x)$
\begin{mathletters}
\label{a14}
\begin{eqnarray}
\int_0^A dxxC^A(x)=&&\int_0^A dz z f^{PN,A}(z)\int_0^1 dttC^N(t)
\label{a14a}\\
\int_0^Adx x [u_v^A+d_v^A+2{\bar u}^A+2{\bar d}^A+2s^A+g^A]_x=&&
\int_0^Adz z f^{PN,A}(z) *
\nonumber\\
\int_0^1 dtt&&[u_v+d_v+2{\bar u}+2{\bar d}+2s+g]_t
\nonumber\\
\approx \int_0^1 dtt&& [u_v+d_v+2{\bar u}+2{\bar d}+2s+g]_t
\label{a14b}
\end{eqnarray}
\end{mathletters}
Again, Eq. (\ref{a14b}) is a special case, related to the momentum sumrule,
which does not exactly carry over to the nuclear case. However, the peaking 
of the normalized $f^{PN,A}$ causes the $z$-integral in Eq. (\ref{a14a}) 
to be very close to 1. In the case of the nuclear momentum sumrule  Eq. 
(\ref{a14b}), the deviations are really minute (see Ref. \onlinecite{rw1}
for an entirely different way to mend the momentum sumrule violation in the
PWIA).

The incompleteness of the above-mentioned Eq. (\ref{a3a}) for $x\lesssim 
0.2$ does not in practice constitute a problem. For one, the conservation
of the number of valence quarks per nucleon is guaranteed by unitarity, 
i.e. the normalization of $f^{PN,A}$. As the momentum sumrule (\ref{a14b})
illustrates, more is requires than unitarity, in case the fact that 
the widths $\Delta x(Q^2)$ of the peaked $f$ are appreciably less than the
support of $x$.  Comparing  the two sides of the expressions in
Eq. (\ref{a14b}) for 4 targets and $Q^2=3.5, 5.0\,$GeV$^2$, we find 
differences of no more than $\approx 1\%$. 

The above sum rules involve pdf in a nucleus and in a nucleon. Various 
sumrules involve moments of nuclear SF and one may compare numerical results 
in the hadronic representation (see for instance Ref. \cite{abb,rtv,arm}) and 
in the partonic one. For instance, Eq. (\ref{a8}) and (\ref{a14b}) hint, that 
there exists a sumrule for $M_{-1}^A \equiv F_2^A/x= \sum a_iq_i^A$. The 
contributions of valence quarks is 5/6, but all other parts diverge (see for 
instance \onlinecite{rob,tow}). Those divergences cancel in differences 
of any pair of those ratios. Thus from Eqs. ({\ref{a8}), ({\ref{a11}) (for 
simplicity we disregard $\delta N/A$ corrections)
\begin{eqnarray}
M^A_{-1}-M^{A'}_{-1}&=&
\int_0^A  \frac {dx}{x} F_2^{A;NI}-\int_0^{A'}\frac {dx}{x}F_2^{A';NI}
\nonumber\\
&=& \frac{5}{18}\bigg [\int_0^A dx (u_v^A+d_v^A)-\int_0^{A'} dx
(u_v^{A'}+d_v^{A'})\bigg ] \approx 0
\label{a15}
\end{eqnarray}
For all $A'\ne A$ (including $A'=1$ with a  $x$-support, maximally different 
from (0,$A$) for a target $A$), the upper integration limits in Eq. 
(\ref{a15}) are unequal. However, for the above-mentioned reason, one may 
neglect the contributions to the integrals in (\ref{a15}) for $x\lesssim 
0.20$ and $x\gtrsim 0.95$. Hence there effectively is an approximate common 
upper limit 
$x_U\approx 1\ll A,A'$, beyond which the difference $(F_2^A-F_2^{A'})$ is 
negligible and the same holds for $x_L\lesssim 0.18$. For the pair D, Fe 
and the chosen three $Q^2$ values, the above difference of the integrals in 
the hadron and pdf representations is $\approx -0.03$ \footnote{ A special 
case occurs for members of an iso-doublet $A=A'$, for which Eq. (\ref{a15}) 
is a generalized Gottfried sum. For a recent discussion of its apparent 
shortcomings, we refer to Ref. \cite{guz}.}.
          
Of special interest is the zeroth moment of nuclear SF, which is related 
to the momentum sumrule. Using Eq. (\ref{a3a})  
\begin{eqnarray}
M^A_0=\int_0^A dx F_2^{A;NI}(x)
&=&\int_0^A dx\int_x^A dz f^{PN,A}(z)F_2^{\langle N\rangle}(x/z)
\nonumber\\
&=&M_0^{\langle N\rangle}\int_0^Adzzf^{PN,A}(z) \approx M_0^{\langle N\rangle}
\label{a16}
\end{eqnarray}
The same moment of iso-singlet NI parts in the pdf representation is from
Eqs. (\ref{a8}), (\ref{a11}) and (\ref{a12b}) seen to be 
\begin{mathletters}
\label{a17}
\begin{eqnarray}
M_0^A=\int_0^A dx F_2^{A;NI}(x)&=&
\frac{5}{18}\int_0^A dxx \bigg [u_v^A+d_v^A+
2({\bar u}^A+{\bar d}^A)+\frac{4}{5}s^A \bigg ]_x 
\nonumber\\
&&\approx \frac{5}{18}\bigg
[1-\int_0^A dx x\bigg (\frac{6}{5}s^A +g^A \bigg )_x\bigg ]
   \label{a17a}\\
&\approx & \frac{5}{18}\bigg [1-\int_0^1 dx x\bigg (\frac{6}{5}s(x) +g(x) \bigg )
\bigg ]=\int_0^1 dx F_2^{\langle N \rangle}(x) 
   \label{a17b}
\end{eqnarray}
\end{mathletters}
NE parts are small for the considered $Q^2$. When included, the 
normalization of  $f^{PN,A}$ guarantees $M_0^{A,NE}=M_0^{N,NE}$, i.e. NE
parts of $M_0^A$ are also $A$-independent. 

In Figs. 2a-2e we show differences of valence, sea quark and gluon
distributions functions in a nucleus and for the $p$. We choose 5 targets 
and display results only for $Q^2=5\,$ GeV$^2$, because for the $Q^2$ range
considered there is hardly any $Q^2$-dependence. Differences increase with 
increasing mass number, and change sign at roughly $x=0.2$ and 0.8. 

To the extent that nuclear sea and gluon distributions are close to the
nucleonic ones, Eq. (\ref{a17b}) shows that in the pdf representation, 
$M_0^A$ is practically $A$-independent. Emphasis is on the standard
$\approx 50\%$ reduction of the nucleon valence contributions 5/18  due to 
gluons (see for instance \onlinecite{rob}), which carries over to nuclear df.

It is not feasible to directly verify the hadronic result (\ref{a16}), which 
requires $F_2^A$ to be known over the entire relevant $x$-range. Present 
data are for $x\gtrsim 0.3$ for the lowest $Q^2$ (for which Eq. (\ref{a16}) 
is not accurate) and $x\gtrsim (0.5-0.6)$ for medium $Q^2$. For both 
the missing information for $x\lesssim 0.5$ contains the major contribution 
to $M_0^A$. We thus take recourse to nuclear SF, computed from Eq. 
(\ref{a3a}). Wherever data are available for the DIS $x$-range, those 
appear to agree very well with computed ones.
  
Table I displays the zeroth moments $M_0^A$, $\,\,A=\{$D, He, C, Fe$\}$, 
computed in the hadronic and the partonic representations (Eq. (\ref{a16}), 
respectively (\ref{a17a}), (\ref{a17b})). Comparison of columns 2 and 3
clearly shows that the shape $f^{PN,A}$ effectively cuts the upper limit 
of long-range integrals at $x_U\approx 1.0$.

The lowest moments $M_0^A(Q^2)$ are seen to be dependent on $A$ in a weak 
and not smooth fashion. Going from D to Fe, those moments are 3-4, 
respectively 8$\%$ smaller than the same for the averaged nucleon 
$M_0^{\langle N \rangle}(Q^2)$: D and He clearly do not follow the  
smooth behavior of all other nuclei. One also notices that $M_0^A$ are
weakly descending functions of $Q^2$.

At this point we return to a verification of the result (\ref{a16}).
In spite of the lacking data on $F_2^A$ for small-$x$, it is possible to
reach an indirect, approximate experimental verification (\ref{a16}) in the
hadronic representation. The method exploits knowledge on $F_k^A$ for
small $x$ and the smoothness of the same in the region $x\gtrsim x_m$,
$x_m$  being the above-mentioned, smallest $x$ measured \cite{arr1,nicu}.
One then interpolates nuclear SFs in the intermediate region, where data are
missing \cite{rtn} and subsequently approximate calculates the lowest
moments $M_0^A$. In spite of the fact, that the $'$missing$'$ region
contributes the major part of the integrals (\ref{a16}), the moments turn
out to be surprisingly close to the ones computed above. Forthcoming data
from JLab experiment E03-103 \cite{arrgas} will enable a sharpening of
the above method.

The approximate $A$-independence of the above zeroth moments of any pair of SF 
entails the same for the differences
\begin{eqnarray}
M^A_0-M_0^{A'}=\int_0^A dx F_2^{A;NI}-\int_0^{A'}dx F_2^{A';NI}\approx
\int_{x_0}^{x_U} dx [F_2^{A;NI}- F_2^{A';NI}]\approx 0
\label{a18}
\end{eqnarray}
The vanishing of the difference is attributed to about equal effective 
$x$-ranges ($x_0,x_U)\approx (0.15, 0.85)$, which replace actual unequal 
supports \cite{arneo1}. Eq. (\ref{a18}) implies that in the above common 
interval, the difference of two SF has to change sign at least once, 
or in different terms: the generalized EMC ratios $\mu^{A,A'}(x,Q^2)=
F_2^A(x,Q^2)/F_2^{A'}(x,Q^2)$ passes the value 1 in the above $x$-interval, 
as is 
indeed observed for all $A,Q^2$ (see Ref.\cite{rtvemc} for a discussion in 
an entirely different context).

The simplest cause for approximate $A$-independence of $M_0^A$ would be the 
same for $F_2^A$, but that appears not to be the case. In the dominant 
classical region $x\lesssim 0.90$, differences $F_2^A-F_2^{A'}$ in both 
representations grow with $x$ beyond $\approx 0.18$ and may become as 
large as 50-60 $\%$, which is far larger than the spread in $M_0^A$ (cf. 
Table I). We shall shortly return to this point.

\section{ EMC ratios in the parton distribution functions representation.}

We have computed $F_2^{A,NI}$ in the pdf representations, using Eq. 
(\ref{a8}), (\ref{a10}) and (\ref{a11}). To those we added the 
NE components (\ref{a5b}), which are only relevant for $|1-x|\lesssim 0.05$. 
The total EMC ratios $F_2^A/F_2^D$ 
are then compared  with recently determined counterparts in the hadron 
representation \cite{rtvemc}.  For the range $0.2 \lesssim x\lesssim 1.2$ 
those, and avaibale data for He, Fe and Au \cite{gomez,bodek,amad}, are 
shown in Figs. 4a,b for $Q^2=3.5, 5.0\,$GeV$^2$.

Up to $x\lesssim 0.65$ there is close agreement between the computed ratios. 
Beyond that point deviations set in, which grow while pdf $\gg$ hadr. The 
latter situation is reversed for $x\gtrsim 0.90$. Both 
representations overestimate the relative maxima in $\mu^A$ around $x=0.9$, 
but over the entire $x$-range, the hadronic results are closer to the data 
than those for the pdf. We attribute this to inferior $F_2^n$ input for 
larger $x$, which propagates into $F_2^A$ through the use of Eq. (\ref{a3a}). 
Also of interest are the slightly lower pdf results for $x\lesssim 0.65$ 
and the much higher ones beyond up to $x\approx 0.9$. 

The apparent insensitivity of EMC ratios $\mu^A$ to the representation, 
in spite of the large differences in the participating $F_2^A$, combined 
with points a), b) in Section III suggest the following: Irrespective
of the cause of the dependence of nuclear SF on $A$ and/or representation, 
the approximate independence of the zeroth moments (\ref{a16}) on both, 
forces the differences in $F_2^A$ in the regions $x \lesssim 0.18$ and 
$0.18\lesssim x\lesssim 0.90$ to be nearly balanced. Consequently, if 
EMC ratios in one area are in some way ordered in $A$, that ordering is 
inverted in the second one. It almost seems that deviations of EMC ratios 
from 1  can be generated by an integral-preserving, affine transformation
with $x\approx 0.18, 0,90$ as fixed-points, and having in particular for 
$A\lesssim 12$, a characteristic $A$-dependence, and having  $x\approx 0.18, 
0,90$ as fixed-points.

Finally, out of sheer curiosity we followed the pdf results down to 
$x=10^{-5}$. Results for $F_2^A$  hardly change from their $A$-independent 
values around $x=0.15$, causing  EMC ratios to stay close to 1 for
decreasing $x$. The above is actually observed down to $x\approx 10^{-3}$. 
Only for the smallest $x$ do screening effects deplete df and cause $\mu^A$ 
to slowly reach values $\approx 0.6-0.7$.

\section{Comparison, conclusions.}

In this note we defined df for partons in a nucleus, using for the 
latter a relation between nucleonic and nuclear hadron SF without any 
adjustable parameter. We showed that either exactly or very closely, those 
respect basic sumrules. Observables such as $F_2^A$ may be expressed 
in either representation, and by construction produce in principle the same 
$F_2^A$ for identical input $F_2^{p,n}$. 

We have shown that this is not the case in practice, specifically not for 
$F_2^n$. The pdf choice rests on the $'$primitive$'$ approximation $F_2^n=
2F_2^D-F_2^p$, which is satisfactory for $x\lesssim 0.3$, but deteriorates  
with increasing $x$, while a well-founded extraction method has been used 
in the hadronic representation. Consequently EMC ratios, computed in the 
two versions, practically coincide for $x\lesssim 0.65$. For large $x$  
deviations appear, which reflect the similarly deviating $F_2^n$ from a 
better founded extracted function: the hadronic representation of EMC 
ratios produces the better fits to the data.

The above, and the fact that deviations of EMC ratios from 1 appear in
distinct areas to be balanced for all $A$, seems linked to the lowest moment 
$M_0^A$ of $F_2^A$: those are for any $Q^2$ practically independent of $A$.
  
We continued pdf calculations down to the smallest $x$, which is the region
were the above criticism does not hold. However, there Eq. (\ref{a10})
misses primarily (anti-)screening effects. Nevertheless, down to 
$x\approx 10^{-3}$ the agreement with data persists, but the pdf results 
cannot describe anti-screening depletion of df in $\mu^A$ for the smallest
measured $x$.

Our almost natural choice of distribution functions of partons in a nucleus 
is clearly one out of many possible ones, and we shall mention a few 
suggested alternatives. For instance Eskola $et\,al$ address participating 
nuclear SF in EMC ratios, which are generated from parametrized input for 
a reference $Q_0^2$, and which are subsequently evolved  to the desired 
$Q^2$ \cite{esk}. Parameters are constrained, for instance by fixing the 
average position of minima of EMC ratios. Data determine those parameters. 

Next we mention Kumano and co-workers who, in spite of different $x$-support 
for $F^A$ and $F^N$, assume a  linear relation between df 
for partons in a nucleus and in a nucleon \cite{kum}. The species-dependent, 
relating weight functions $w_i(x,Z,A)$ contain parameters for a scale 
$Q_0^2$ and the resulting df of partons in a nucleus are again evolved to 
any desired $Q^2$, ultimately producing parametrized nuclear SF and EMC 
ratios. A large number of adjustable parameters leads to fits from the 
smallest $x$ up to $x\le 1$. With the connecting weight function having 
no meaning beyond $x=1$, the interesting region $x\ge 1$ is out of reach in 
that approach. A more serious drawback of the method may be the lack of 
physical meaning of the weight functions $w_i$ and its parameters.

Finally we discuss approaches, where df of partons in a nucleus 
are those for a nucleon bound in scalar and vector mean fields 
\cite{mineo,stef}, which couple to quarks in a nucleon \cite{guich1}. 
Off-hand, the above and our phenomenological approach seem to have little 
in common, but this is actually not the case, and it is instructive to 
trace the connection. We recall that the original proposal Eq. (\ref{a3a}) 
was inspired by a model, where valence quarks in a nucleus cluster in 
bags \cite{gr}. Total interactions between quarks in two different bags 
have there been replaced by phenomenological $NN$ forces, acting on the 
centers of those bags, thus replacing quark dynamics by those for 
hadrons. The above treatment of a nucleus partially calls back
quark degrees of freedom, and are from Eq. (\ref{a9}) seen to mix the 
latter with hadronic degrees of freedom through the SF $f^{PN,A}$. 

In contrast, the hybrid meson-quark coupling model, where the interactions 
of a single quark in a given nucleon with all quarks in the remaining $A-1$ 
bags are  replaced by mean fields. As expected from its intermediate 
position, it is possible to sum in that model the above meson-quark 
interactions over the valence quarks in a bag, and to construct a $NN$ 
interaction with many-body components, mediated by the same mean boson 
fields. This has recently been shown to be possible \cite{guich2}. 

The above has been  applied to nuclear matter \cite{mineo,stef} and it is 
clearly of interest to see applications of finite nuclei. A first example
is a treatment of $^3$He in the PWIA \cite{ss} and one should look forward
to results for higher $A$.

\begin{figure}[p]
\includegraphics[bb=-150 440 567 400,angle=-90,scale=.70]{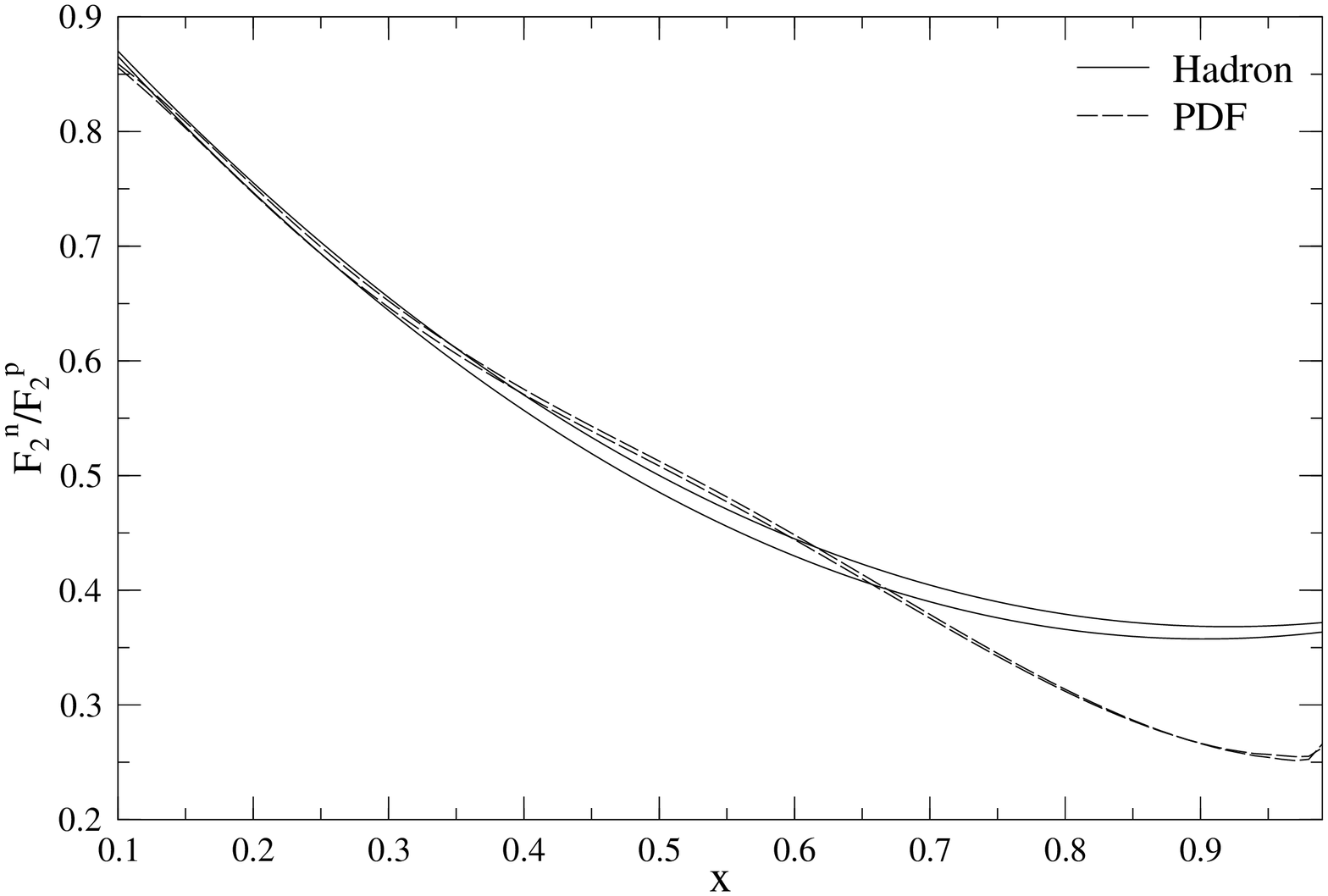}
\caption{  The ratio $F_2^n(x,Q^2)/F_2^p(x,Q^2)$ in the hadronic (drawn
lines) and the pdf representation (dashes); Upper and lower curves
correspond to $Q^2=3.5, 5.0\,$GeV$^2$.}
\end{figure}

\begin{figure}[p]
\includegraphics[angle=0,scale=.70]{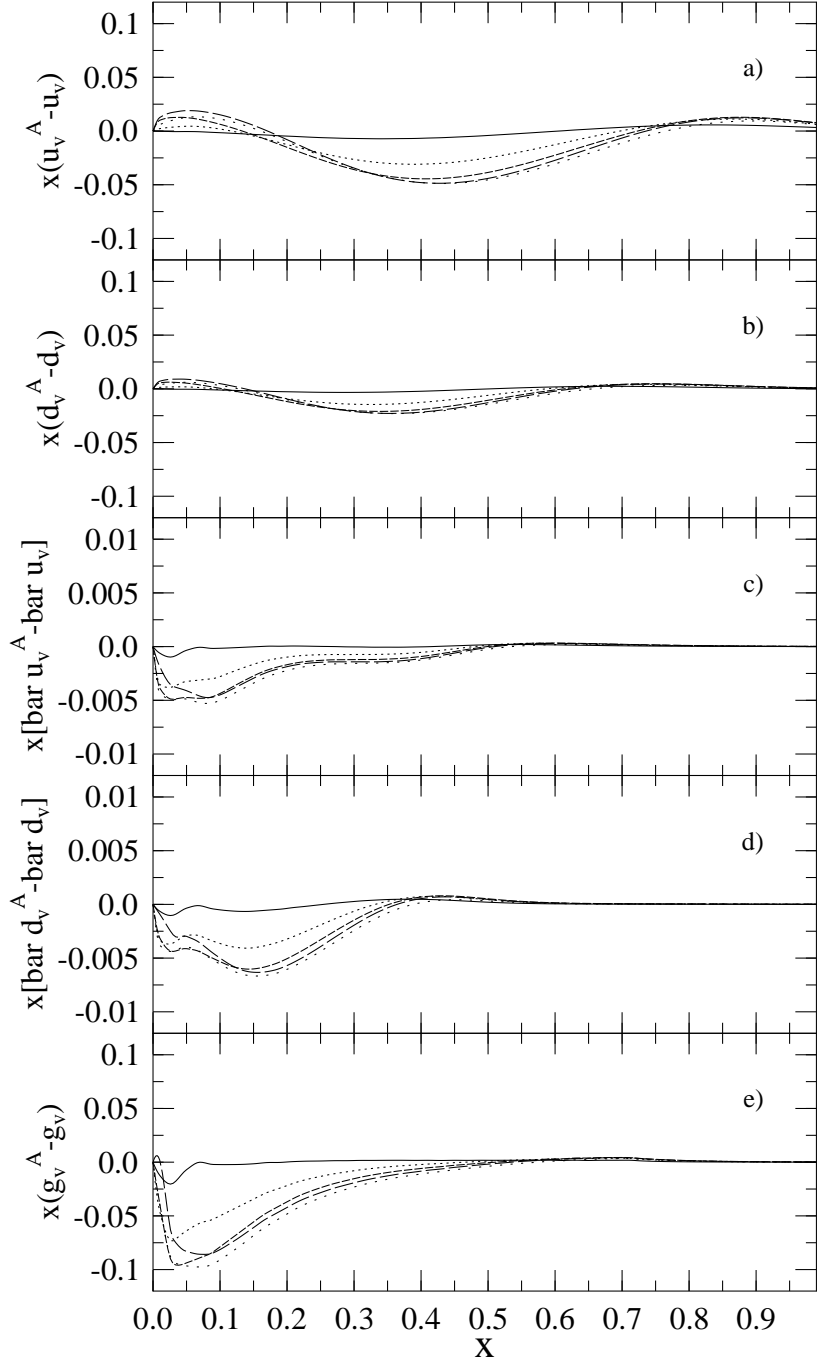}
\caption{  Differences $x(q^A-q)$ for D (drawn line), $^4$He (dots), C (spaced
dots), Fe (short dashes) and Au (long dashes) and a $p$; $Q^2=5\,$GeV$^2$.
a): $q=u_v$; b) $q=d_v$; c) $q={\bar u}$; d) $q={\bar d}$; e) $q=g$.}
\end{figure}                          

\begin{figure}[p]                                                
\includegraphics[bb=-150 440 567 400,angle=-90,scale=.70]{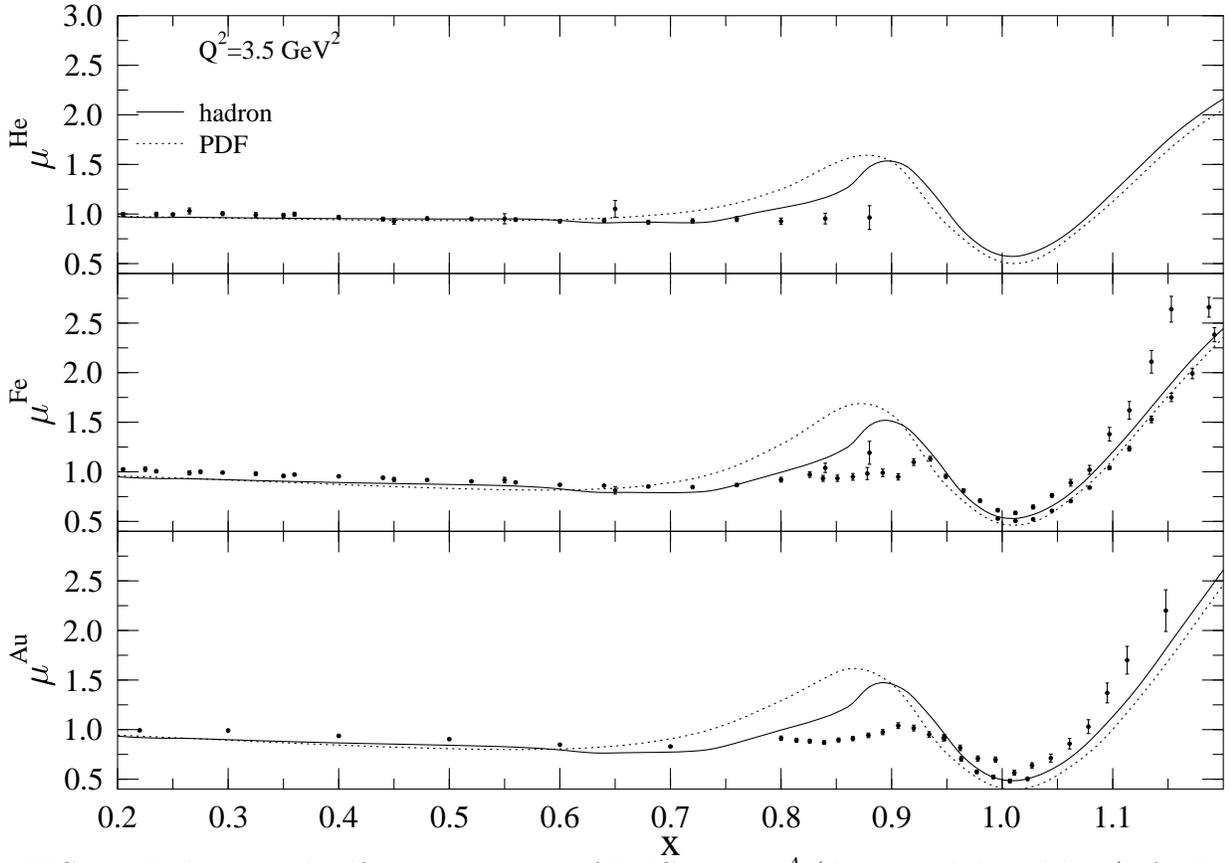}
\caption{  Hadronic and pdf representations of EMC ratios $\mu^A$ (drawn and
dotted lines), A=He, C and Fe  for $0.2 \lesssim x\lesssim 1,2\,\,$, 
$Q^2=3.5\,$GeV$^2$. Data are from \cite{gomez,bodek,amad}.}
\end{figure}

\begin{figure}[p]
\includegraphics[bb=-150 440 567 400,angle=-90,scale=.70]{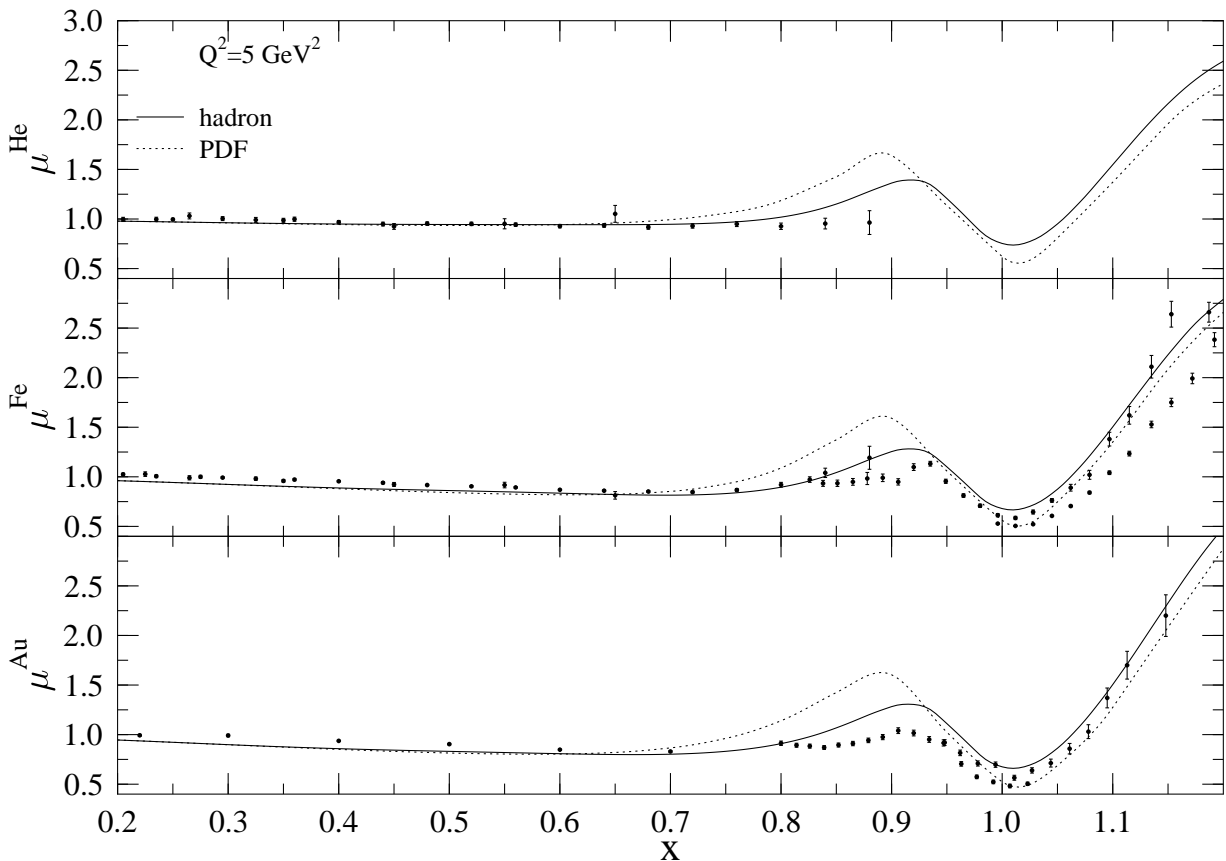}
\caption{  Same as Fig. 3 for  $Q^2=5\,$GeV$^2$.}
\end{figure}

\begin{table}
\caption {Zeroth moments of $F_2^A$ in the hadron (h) and a parton (p)
representation, with upper limits $A$ approximated by 1 and 2.} 

\begin{tabular}{|c|c|c|}
\hline
target   &$\int_0^1 dx F_2^A(x,Q^2)$ & $\int_0^2 dx F_2^A(x,Q^2)$ \\
\hline
D~~~~~~~h  & 0.1492, 0.1484, 0.1409~~~~~~~~~~~~~~~ & 
             0.1493, 0.1484, 0.1409~~~~~~~~~~~~~~~ \\
~~~~~~~~p  & 0.1505, 0.1470, 0.1413~~~~~~~~~~~~~~~ & 
             0.1506, 0.1470, 0.1413~~~~~~~~~~~~~~~ \\
\hline
$^4$He~~h  & 0.1455, 0.1450, 0.1378~~~~~~~~~~~~~~ & 
             0.1459, 0.1453, 0.1379~~~~~~~~~~~~~~ \\
~~~~~~~~p  & 0.1464, 0.1433, 0.1378~~~~~~~~~~~~~~ & 
             0.1467, 0.1435, 0.1379~~~~~~~~~~~~~~ \\
\hline
C~~~~~~~h  & 0.1434, 0.1434, 0.1370~~~~~~~~~~~~~~ & 
             0.1440, 0.1439, 0.1372~~~~~~~~~~~~~~ \\     
~~~~~~~~p  & 0.1430, 0.1403, 0.1353~~~~~~~~~~~~~~ & 
             0.1434, 0.1408, 0.1383~~~~~~~~~~~~~~ \\
\hline
Fe~~~~~~h  & 0.1402, 0.1403, 0.1338~~~~~~~~~~~~~~ & 
             0.1406, 0.1406, 0.1339~~~~~~~~~~~~~~ \\     
~~~~~~~~p  & 0.1447, 0.1403, 0.1353~~~~~~~~~~~~~~ & 
             0.1433, 0.1405, 0.1351~~~~~~~~~~~~~~ \\
\hline
$\langle N\rangle$ ~~h& 0.1554, 0.1510, 0.1438~~~~~~~~~~~~~~ &           \\   
~~~~~~~p              & 0.1510, 0.1475, 0.1420~~~~~~~~~~~~~~ &           \\
\hline
\end{tabular}

\label{Table I}
\end{table} 
\end{document}